 \documentclass{elsart}
\usepackage{epsfig}
\usepackage{caption}

\usepackage{lineno}
\usepackage{setspace}

\linenumbers

\begin{document}

  \begin{frontmatter}
    \title{Measurements of radio propagation in rock salt for the detection of high-energy neutrinos}
    \author[ucl]{Amy Connolly}, 
\ead{amyc@hep.ucl.ac.uk}
    \author[ucla]{Abigail Goodhue}, 
    \author[uh]{Christian Miki}, 
    \author[ucl]{Ryan Nichol}, 
    \author[ucla]{David Saltzberg}
    \address[ucl]{Department of Physics and Astronomy, University College London, Gower Street, London  WC1E 6BT  UK}
    \address[ucla]{Department of Physics and Astronomy, UCLA, 475 Portola Plaza, Mailstop~154705, Los Angeles, CA 90095-1547  USA}
    \address[uh]{Department of Physics and Astronomy, University of Hawaii, 2505 Correa Rd., Honolulu, HI  96822  USA}

    \begin{abstract}
We present measurements of the transmission of
radio/microwave pulses through salt in the Cote Blanche salt mine operated
by the North American Salt Company in
St.~Mary Parish, Louisiana.  These results are from 
data taken 
in the southwestern region of the 1500~ft. (457~m) deep level of the mine 
on our third and most recent visit to the mine.
We transmitted and received 
a fast, high-power, broadband pulse from within three vertical boreholes
that were drilled to depths of 100~ft. (30~m) and 200~ft. below the 1500~ft. level
using three different pairs of dipole antennas whose 
bandwidths span 125~to~900~MHz.  By measuring the relative strength of
the received pulses between boreholes with separations of 50~m and 169~m, 
we deduce the attenuation of the signal attributed to the salt medium.   
We fit the frequency dependence of the attenuation to a power law 
and find the best fit field attenuation lengths to be 
$93 \pm 7$~m at 150~MHz,
$63 \pm 3$~m at 300~MHz,
and $36 \pm 2$~m at 800~MHz.
This is the most precise measurement of
radio attenuation in a natural salt formation to date.
We assess the implications of this measurement for a future neutrino detector in
salt.

    \end{abstract}
   
  \end{frontmatter}

\setstretch{2}  
  \section{Introduction}
  The observation of cosmic rays with energy higher than the Greisen-Zatsepin-Kuzmin~(GZK) 
  cutoff at $\sim~10^{19.5}$~eV~\cite{gzk} implies a corresponding flux of ultra high-energy (UHE) neutrinos, 
  with energy in the $10^{17}-10^{19}$~eV range~\cite{ess}.  These secondary neutrinos 
  are created via photomeson 
  production of cosmic rays on the 2.7~K cosmic microwave background.  Detection of these neutrinos 
  would provide unique information about the origin of primary cosmic rays and the nature 
  of their sources.
  
  Although predictions for the flux of UHE
  neutrinos differ by orders of magnitude, a reasonable set of parameters
  puts the rate of UHE neutrinos at the level of 10/km$^2$/century~\cite{ess}.
  A detector volume of hundreds of cubic kilometers 
  water equivalent is required to detect a significant 
  number of neutrinos in this energy regime.  Optical techniques are widely used in 
  neutrino detectors, but the volume that can be instrumented is 
  constrained by attenuation lengths of tens of meters
  at optical frequencies in detector media (such as ice), so 
  optical detectors have limited sensitivity to rare high-energy events.
  
 G.~Askaryan first predicted coherent radio emission from ultra 
  high-energy showers~\cite{askaryan}, and the effect has been confirmed in 
  accelerator beam tests~\cite{sandsaltice}. 
 The energy of the coherent radio emission that results from the development 
  of a negative charge excess in the shower depends quadratically on shower energy, and for 
  showers with energy greater than $10^{16}$~eV, dominates 
  the emitted Cherenkov power spectrum.  

 Askaryan also proposed a few materials as detection media
 that occur naturally in very large volumes and are expected to have 
 long attenuation lengths in the radio regime, including ice and salt. 
 Radio attenuation 
 lengths longer 
 than 1~km have been measured in ice near the South Pole~\cite{southpoleice}.

  Formations of salt rock could therefore be a viable detector for high-energy neutrinos if 
  attenuation in the radio regime is indeed low.  
  Domes of rock salt occur naturally in many parts of the world, including 
  the Gulf Coast region of the United States.  In these formations, 
the salt originates from
  dried ocean beds which have been buoyed upward due to geological forces
  through a process called diapirism.  Through this process, the salt
  becomes very pure as impurities are extruded.
  Such salt domes, with typical dimensions of several square kilometers by several kilometers deep, 
  are thought to be good candidates for a neutrino detector.  
  For a more detailed discussion of the dielectric properties of salt and the application to neutrino 
  detection, see Reference~\cite{hockley}.

  \section{Previous Measurements}
  There are two classifications of measurements that have been made of the radio properties 
  of rock salt in salt domes.  Ground Penetrating Radar~(GPR) was used in the 1960's and 
  1970's to determine the size and structure of salt domes.  This technique is based on sending 
  a radar signal into the salt and 
  looking for reflections from interfaces in the salt structure.  One can calculate the 
  distance of the interface from the time delay of the reflected pulse.  Although the primary purpose 
  of the GPR measurements is to calculate interface distances, one can also extract an attenuation 
  coefficient based on the detected signal voltage, the voltage of the transmitted pulse, and the distance 
  over which the pulse traveled.  Using the GPR technique to extract attenuation is complicated by 
the unknown reflection 
  coefficient and geometry at the surface of reflection, but assuming a coefficient of $1.0$ 
from a flat surface gives a conservative estimate.

  Direct measurements of attenuation in rock salt have also been made.  In 2002, measurements made at 
  Hockley salt mine showed attenuation lengths consistent with being 
longer than 40~m at 150~and~300~MHz~\cite{hockley}.  
  Their measurements were limited by the voltage of the pulser, which limited their
transmission distances in the salt.
We decided
  to follow the techniques of the direct transmission measurements but used a high voltage pulser 
  so that we could transmit through longer distances of salt.  
  
  \section{Cote Blanche Salt Mine}
  We made measurements of the dielectric properties of salt in the mine located in the Cote Blanche 
  salt dome in St.~Mary Parish, Louisiana in August 2007.  
  We chose this dome because GPR measurements that were made in the mine
  suggested very low radio attenuation, including observations of reflections over the longest 
  distance of any mine measured~\cite{s-unterberger}.  
  
  The Cote Blanche dome is one of five 
  salt domes in the area.  The salt dome extends from approximately 90~m below the 
  surface to 4270~m below the surface.  At a depth of 1100~ft.~(335~m), the salt 
  extends 1700~m east to west and 2100~m north to south.  At 2000~ft.~(610~m) deep, 
  the horizontal cross section has nearly doubled in size.   The total volume of 
  salt in the dome is estimated to be 28-30~km$^3$~\cite{halbouty, lock,cherryelog, davidw}.
  
  The dome has been actively mined since~1965 using the 
  conventional ``room and pillar'' method.  The mine consists of a grid of 30~ft.~(9~m) 
  wide by 30~ft.~(9~m) high drifts (corridors) spaced by 100~ft. 
by 100~ft. (30~m by 30~m)
  pillars of salt, and has three levels at depths of 1100~ft., 1300~ft., 
  and 1500~ft.~(335~m, 396~m, and 457~m).  Each level covers 
  several square kilometers.  Current mining operations are on the 1500~ft. deep level.  
  
  The measurements described in this paper are from the third trip that we made to the Cote 
  Blanche mine.  In May 2005, on our first visit to the site, with 
  M. Cherry and J. Marsh from Louisiana State University, we 
  established the viability of the experimental setup
  and measured field attenuation lengths along a corridor (within 10~meters of a wall) on the
1300~ft. level of the mine, 
of approximately
10-15~m at 150~MHz and 300~MHz~\cite{firstvisit}.    
  In September 2006, on a return visit to the mine, we transmitted and 
  received signals horizontally across a 
  single pillar of salt, both with antennas placed against the walls 
  and with antennas inserted 4~m into the ceiling in shallow boreholes.  
  We measured attenuation lengths 
  of $24.6\pm2.2$~m in the frequency range 50-150~MHz, 
  $22.2\pm1.8$~m at 150-250~MHz, 
  and $20.5\pm1.5$~m 
  at 250-350~MHz.  We also measured an average index of 
  refraction of $n=2.4\pm0.1$~\cite{secondvisit}.  

  The attenuation lengths that we 
  measured were inconsistent with previous GPR studies of the Cote Blanche 
  mine (see Section~\ref{sec:gpr}).  
  The relatively short attenuation lengths that we measured through the walls of the 
  pillar could be due to the method used to mine the salt. 
  The miners carve out corridors in the salt by cutting a 
  horizontal slice out from under the wall and then blasting the 
  section above the floor so that the salt can be removed, and
  then proceed to a new section of salt further along the corridor.
  The data from our first two trips were consistent with a model of
  very lossy salt (approximately 10~m attenuation length) 
  in the region closest
  to the walls of the pillar  
  and longer attenuation lengths (quoted above) 
  in salt more than about 10~m from the wall.  
  
  \section{Method}
  
  \subsection{Experimental Setup}
\begin{figure}
    \epsfig{figure=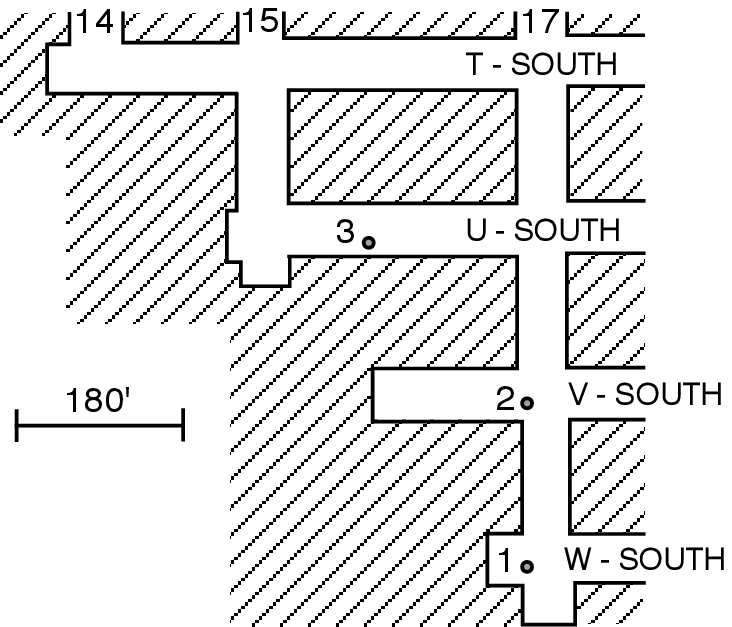, height=4.0in}
    \caption{ }
    \label{fig:saltmap}
  \end{figure}

\begin{figure}
    \epsfig{figure=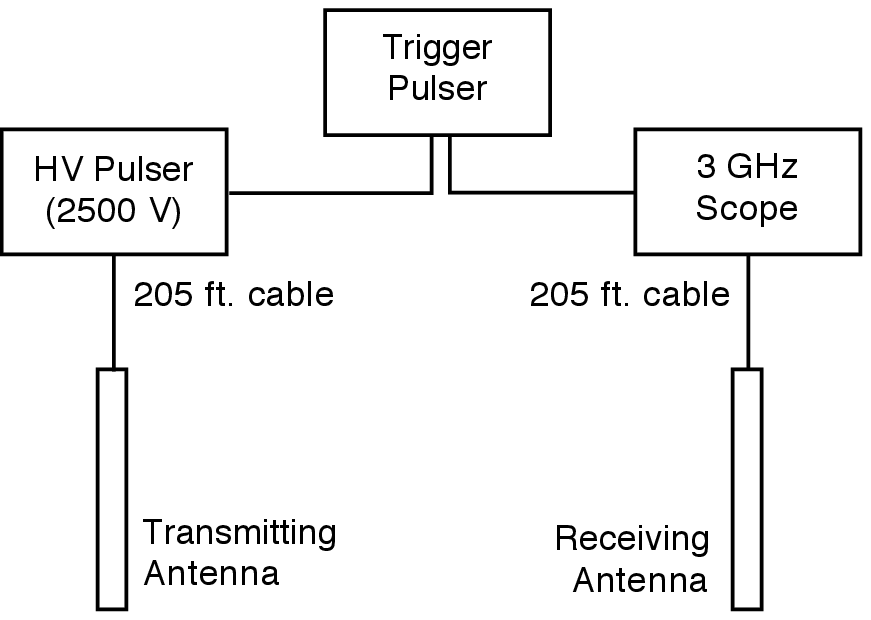, height=3.5in}
    \caption{ }
    \label{fig:systemdiagram}
  \end{figure}

  Figure~\ref{fig:saltmap} shows the region of the 1500~ft. deep level of the mine 
  where we made our measurements, and Figure~\ref{fig:systemdiagram} is 
  a diagram of the experimental setup.  
  The miners drilled three 2.75~inch (7.0~cm) diameter boreholes into the salt of  
  the floor of the 1500~ft. level of the mine.
  Boreholes~1 and~2 were 100~ft. (30~m) deep, while Borehole~3 was 200~ft. (60~m) deep.
  At the first two boreholes, drilling was stopped at 100~ft. once the
  driller encountered methane gas.
  We used a fast, high-power, broadband Pockel's cell 
  pulser model HYPS 
from Grand Applied Physics with a peak voltage of 2500~kV and a 
  10\%-90\%~rise time of 200~ps to generate radio pulses.  
  There was 5~ft. of LMR~240 cable and
  200~ft. of LMR~600 cable leading to the transmitting as well as from the 
  receiving antennas, and the antennas were lowered by hand into the salt.
  
  We made most of our measurements using three pairs 
  of dipole antennas.  We measured the 3~dB points of each antenna when they were  
  in a borehole in the salt to determine the in-band 
  frequencies.  The 3~dB points of the transmission
  of the low frequency (LF) antenna pair (Raven Research RR6335), 
  are 50~to~175~MHz in salt.  For the medium frequency (MF) antennas, which we custom made, 
  the 3~dB points are
  175~and~500~MHz in salt.  The high frequency (HF) pair 
  (Shure Incorporated UA820A) has a transmission 
  band between 550~and~900~MHz in salt.
  
  We used a Tektronix TDS694C oscilloscope to 
  record the received signal pulse.  We used a pulser with synchronized outputs
  to trigger both the the high-power pulser and the oscilloscope.  This allowed
  us to look in a fixed time window for the signal and reduce noise by averaging
  many waveforms.
  
  The transmitting antenna was always lowered into Borehole~1, 
  and the identical receiving antenna was either in Borehole~2, which was 50~m away from the 
  transmitter, or Borehole~3, which was 169~m away from the transmitter.  We took data at 10~ft.~(3~m) 
  incremental depths in the boreholes.  The setup used to measure transmission between 
  Boreholes~1~and~2 was identical to the setup used to 
  measure between Boreholes~1~and~3 
  so that we could make reliable relative measurements without requiring an absolute system 
  calibration.
  Because the transmission band of each antenna was relatively broad, we were able to make 
  measurements between 125~and~900~MHz in salt. 

  \subsection{Attenuation Lengths}

\begin{figure}
    \epsfig{figure=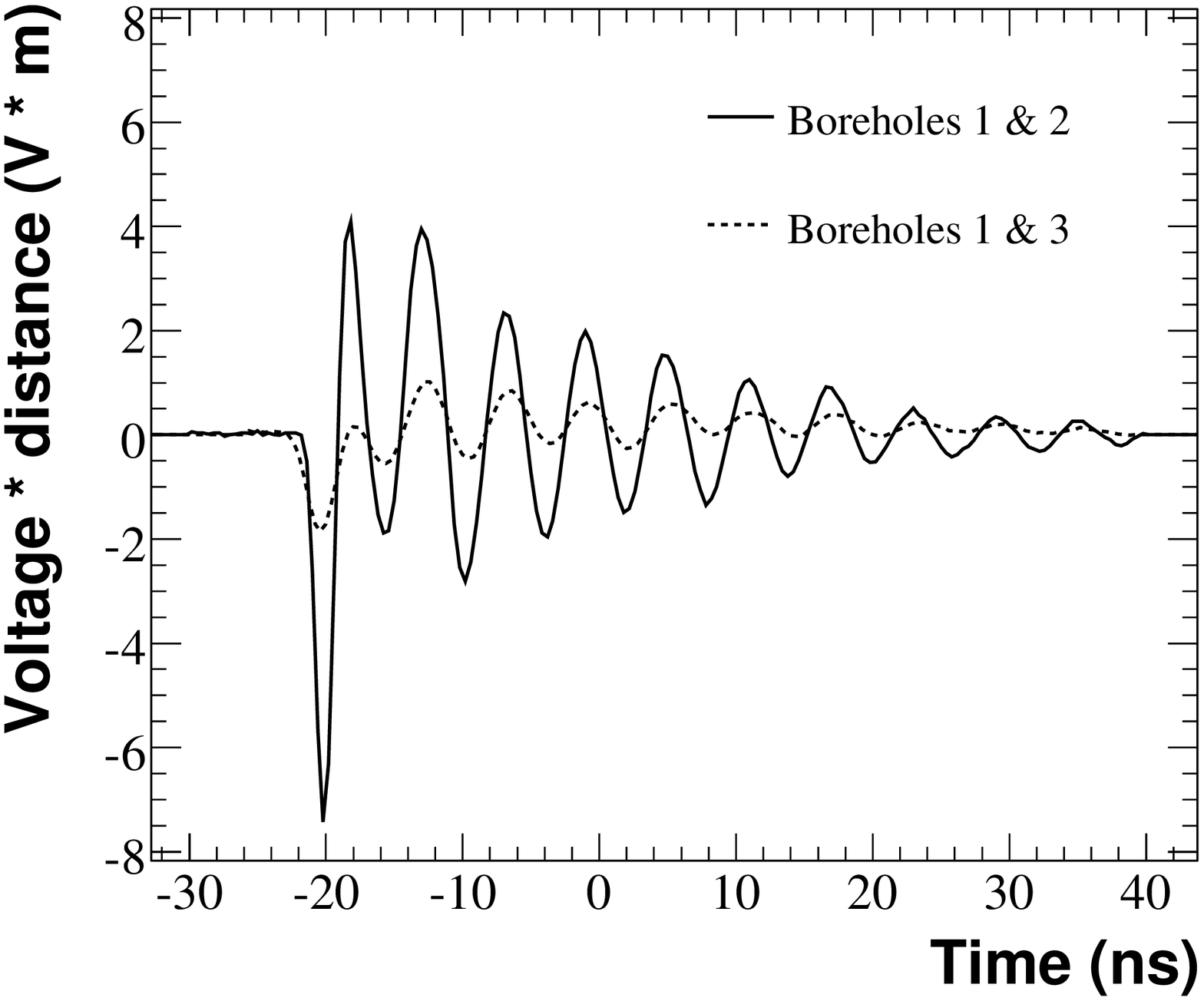, height=4.0in}
    \caption{ }
    \label{fig:timedomain12_13}
  \end{figure}

\begin{figure}
    \epsfig{figure=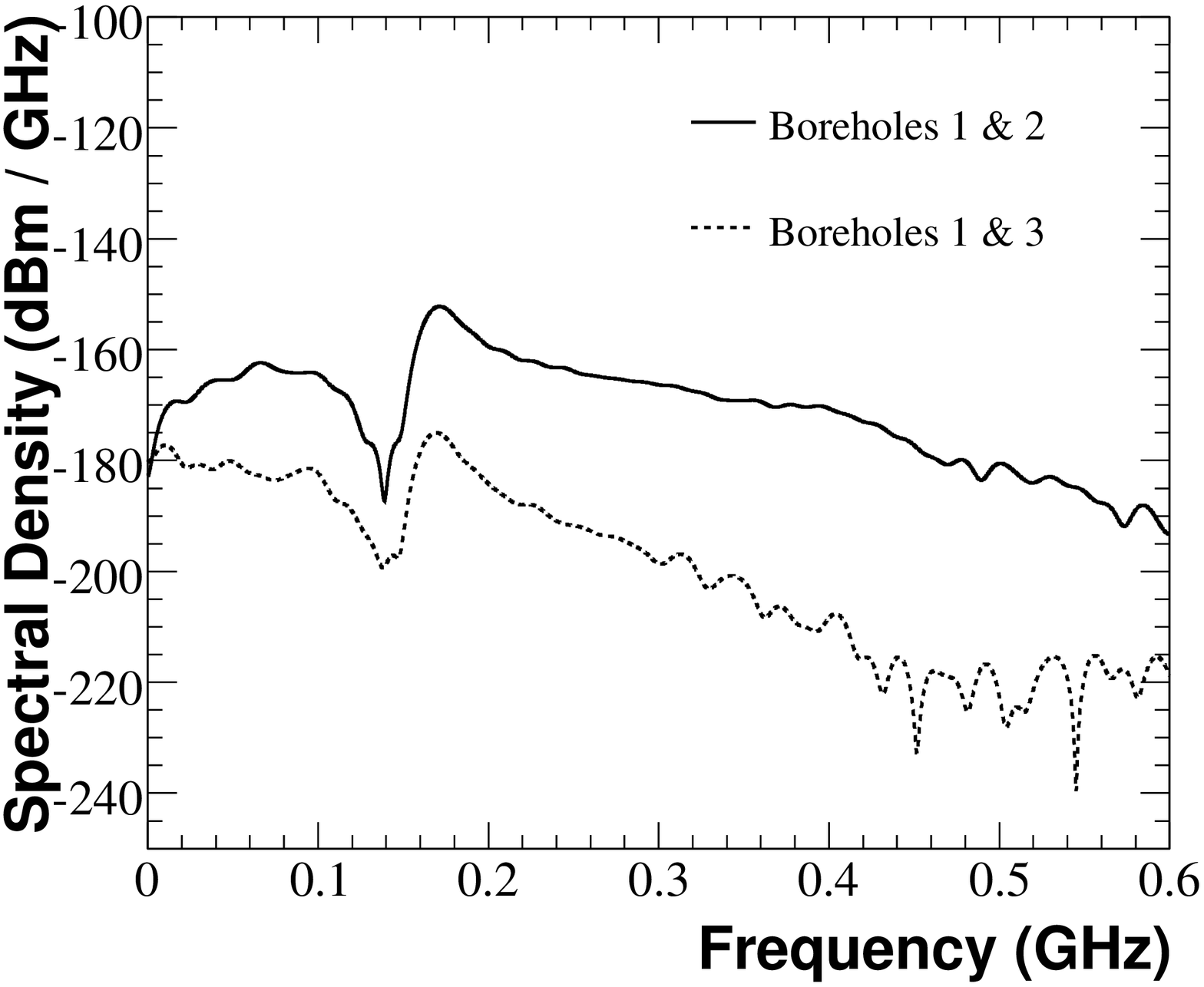, height=4.0in}
    \caption{ }
    \label{fig:freqdomain12_13}
  \end{figure}

  Figure~\ref{fig:timedomain12_13} shows an example waveform 
  of the received pulse at Borehole~2 and Borehole~3 using the MF antennas, 
  and Figure~\ref{fig:freqdomain12_13} 
  shows the Fourier transform of the same waveforms. To calculate the attenuation length, we 
  first cut the recorded waveforms in a time window around the pulse to 
  eliminate any reflections and reduce contributions from noise.  The width 
  of the window is 35~ns for the LF antennas, 30~ns for the MF antennas, 
  and 12~ns for the HF antennas.  


  We expect to record reflections from interfaces within the 
  salt and from the salt-air boundary.  
  We were able to see reflections 
  from the 1500~ft. corridor with the receiver at either borehole.  
  The time window that we use to for the analysis 
  extends no more than 20~ns after 
  the peak of the pulse to ensure that we eliminate 
  the possibility of interference from 
  any reflection off of the corridor for depths of 50~ft. and deeper.

  We sum the 
  total power in frequency bands that are 50~MHz wide 
  for the LF and 
  MF antennas, and 100~MHz wide for the 
  HF antennas which had a larger bandwidth.  
  We subtract the noise contributions in each band.  The noise power is taken from a sample 
  waveform for each antenna type 
  in a time window of the same length as the pulse window 
  but earlier in time than any pulse appears.  The noise subtraction had a small effect on 
  the calculated attenuation length because the signal to noise ratio was high in the band of the 
  antenna.

  We define 
  voltages $V^{i}_{12}$ and $V^{i}_{13}$ to be the square root of the
  power in the $i^{th}$ frequency bin between Boreholes~1 and~2 and Boreholes~1 and~3 which span 
  distances $d_{12}$ and $d_{13}$
  respectively.  Since we used the same system regardless of the location of the 
  receiver, the voltages received in the boreholes from the transmitter are related by:
  \begin{equation}
    \frac{ V^{i}_{13}}{V^{i}_{12}} = \frac{d_{12}}{d_{13}} \cdot \exp\left[-\left(d_{13}-d_{12}\right)/L^{i}_\alpha\right]
  \end{equation}
  where~$L^{i}_\alpha$ is the field attenuation in the $i^{th}$ frequency bin.  
  Inverting this equation gives an expression for the field attenuation length
  in each frequency bin: 
  \begin{equation}
    \label{eq:atten}
    L^{i}_\alpha=(d_{13}-d_{12})/ \ln \left( \frac{d_{12}V^{i}_{12}}{d_{13}V^{i}_{13}} \right)
  \end{equation}
  
  \section{Uncertainties}
  \label{sec:error}
  The main source of systematic uncertainty on the measurement of 
  the field attenuation length at a given depth 
  is due to the position of the antenna within the hole.  
  In previous trips to Cote Blanche, we discovered that a small variation in position of the antenna 
  led to a significant change in the voltage received through the salt~\cite{secondvisit}.  
  We estimate the size of this uncertainty as the root mean square 
  variation between neighboring depths of 
  the peak-to-peak voltage of the recorded waveform at each depth
  (excluding the 10~and~20~ft. depths):
  \begin{equation}
    \delta V = \frac{1}{N} \sqrt{\sum_{i=0}^{N-1} (V_{i}-V_{i+1})^2} 
  \end{equation}
  where $N$ is the number of depths included here. Using this method, 
  we estimate the uncertainty on the voltage measured due to the position of the antenna 
  in the hole to be $24\%$.
  
  
  We also include an uncertainty due to the exact choice of the time window that contains the pulse.  
  We estimate this uncertainty as the root mean square variation 
  of the total power in the pulse as we slide 
  the time window by 1~ns.  When the power at a given frequency is small, 
  this uncertainty dominates (up to 50\% in voltage), but 
  in the frequency band of the antenna, the uncertainty is small (less than 10\% in voltage).
  
  There is also an uncertainty on the distance between the holes.  We 
  measured the distance between Boreholes~1~and~2 
  with a measuring tape, and then used relative 
  timing of received pulses to extrapolate the distance between Boreholes~1~and~3.  
  The uncertainty on these distances include
  a contribution from system timing ($\pm3$~ns), 
  one due to the measurement of the distance between Boreholes~1~and~2 ($\pm1$ ft.), 
  and one  
  due to a depth-dependent potential deviation of the boreholes from vertical ($<\pm3$ ft.).
  The maximum uncertainty on the distance between the transmitting and receiving 
  antennas is always less than $2.1\%$.

\section{Results}

  \subsection{Antenna Transmission}

\begin{figure}
    \epsfig{figure=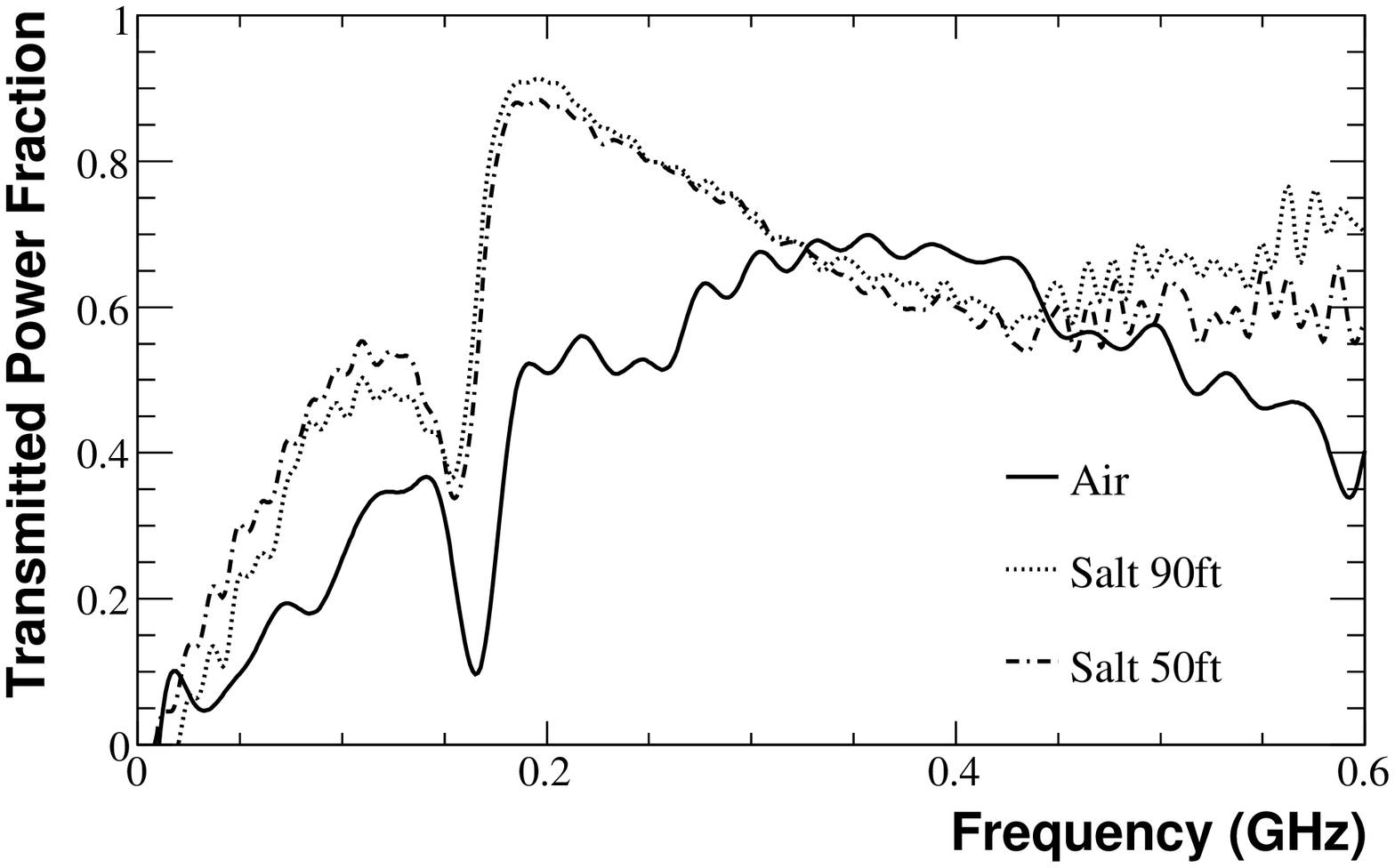, height=3.5in}
    \caption{ }
    \label{fig:s21}
  \end{figure}

  Using the same pulser that was used for the attenuation measurements,
  we measured the fraction of power reflected from each antenna while it
  was in a borehole  
  so that we could deduce the fraction transmitted and frequency dependence. 
  For this S11~measurement, we inserted
  a coupler (Minicircuits ZFDC-20-4) between the pulser and antenna and 
  recorded the reflected signal through the coupled port.
  After measuring the reflection
  from the open cable (with the setup identical but with only the antenna
  removed), the ratio of the power in the 
  two pulses is the fraction reflected from the antenna.  The transmitted
  fraction is deduced by taking the sum of the 
  reflected and transmitted power fractions to be unity. 
  Figure~\ref{fig:s21} shows the transmission 
  of an MF antenna as a function of frequency.  
  The transmission 
  did not change 
  significantly when we changed the depth of the antenna in the hole.   
  
  \subsection{Attenuation Lengths} 
\begin{figure}
    \epsfig{figure=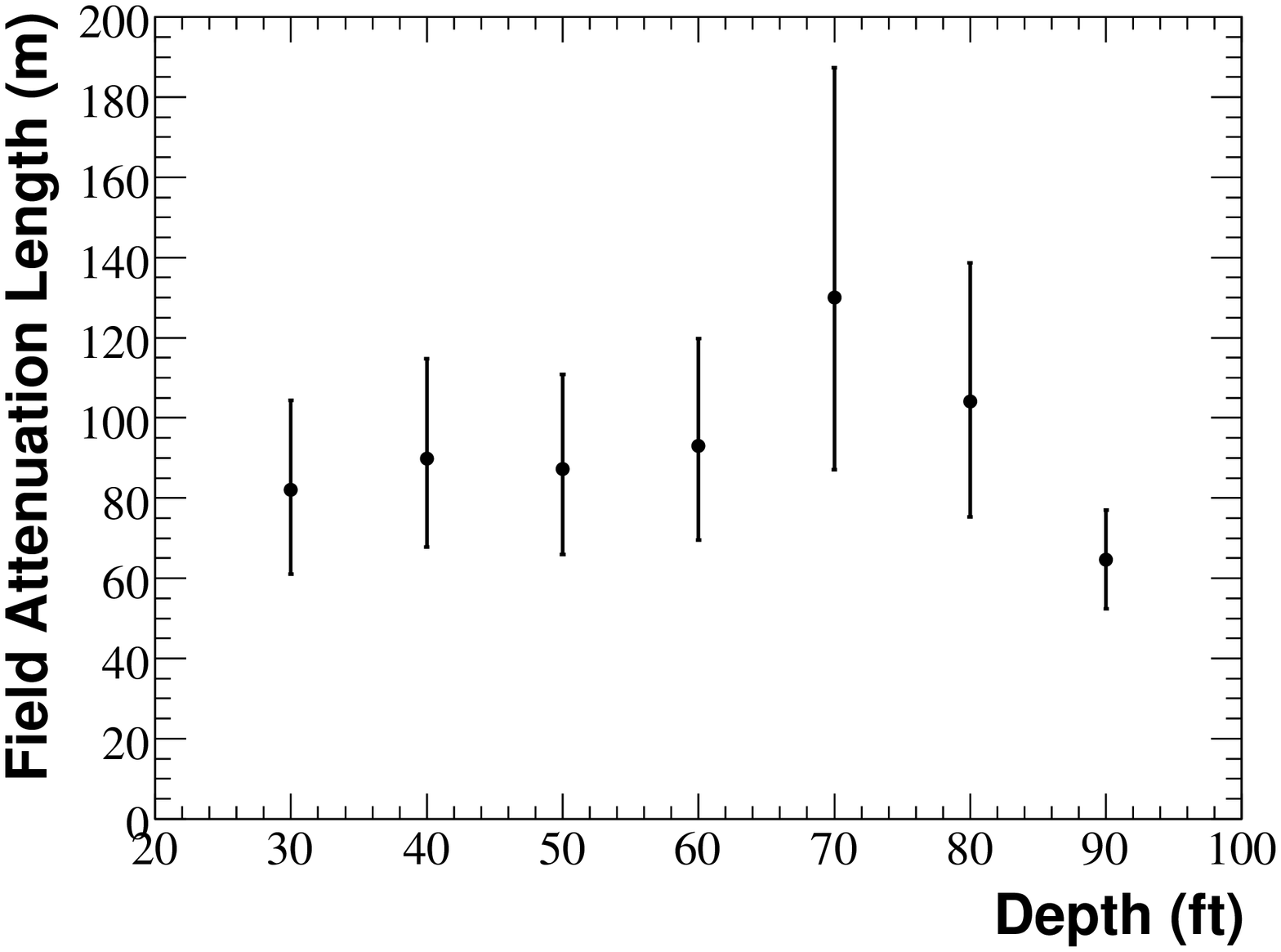, height=4.0in}
    \caption{ } 
    \label{fig:depth}
  \end{figure}

  Figure~\ref{fig:depth} shows the attenuation length in the frequency
  bin centered on 250~MHz as measured with the 
  MF antennas plotted versus depth.      
  We took measurements at 10~and 20~ft. depths, and the values 
  we calculate for the attenuation length at those shallow depths are consistent 
  with attenuation lengths at greater depths; however,
  we only report the results from 30~ft. and deeper
  since the deeper measurements 
  are more likely to be of unfractured, clear salt.  
  At 30 and 40~ft., as much as 1/4 of the power in the
waveforms received at Borehole~3 may be due to
 power from a reflected pulse (based on measurements at greater depths) 
but we still include those points here.
  We do not observe a depth dependence in attenuation length.
  
\begin{figure}
    \epsfig{figure=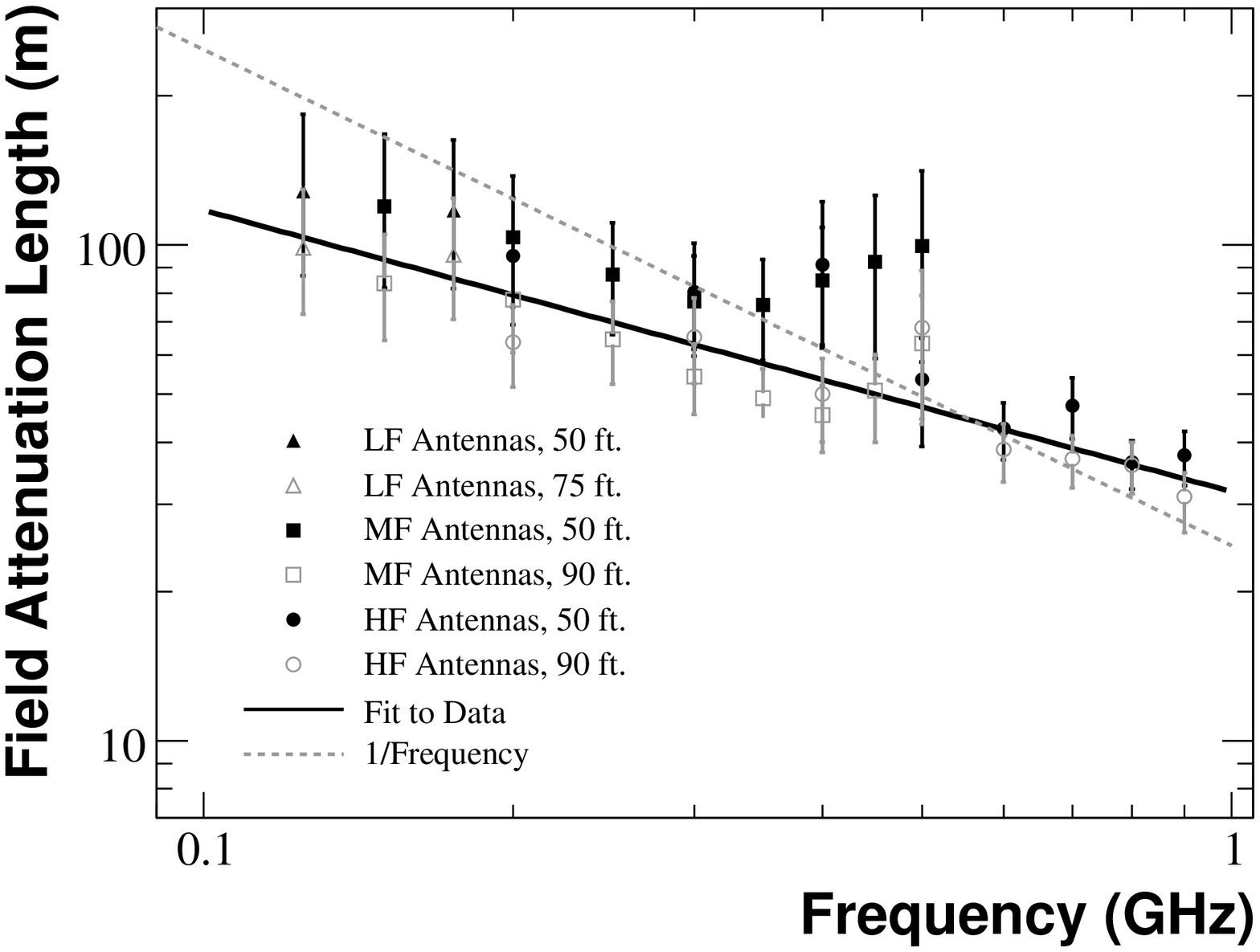, height=4.0in}
    \caption{ }
    \label{fig:freq}
  \end{figure}

  Measured field attenuation length at 50~ft. and 90~ft. depths 
  are shown in Figure~\ref{fig:freq} as a function of 
  frequency.  The LF antennas were not
  easily lowered beyond 75~ft. depth, so we include 
  data from those antennas at
  75~ft. instead of 90~ft.  
  If the salt had a constant loss tangent, we would expect that 
  the attenuation length would decrease with 
  increasing frequency as $\nu^{-1}$, where $\nu$ is the frequency of the radiation~\cite{hockley}.    
  We fit the points in Figure~\ref{fig:freq} to a power law
  of the form:
  \begin{equation}
    L_{\alpha}\left(\nu\right)=a \cdot \left(\frac{\nu}{1~\mathrm{GHz}}\right)^{b}
  \end{equation}
  and the best fit values are $a=32\pm2$~m and $b=-0.57\pm0.06$,
  with a $\chi^2$/dof=25.6/36.
  Although it is clear that
  the attenuation length is falling with increasing frequency, it does not follow the $\nu^{-1}$ 
  expectation, which hints at a non-constant loss tangent, 
  consistent with measurements at the Hockley 
  mine~\cite{hockley}.  In Table~\ref{tab:attnlengths}, we list the measured
attenuation lengths at a few frequencies of interest based
on the best fit values from this fit.
  
The uncertainties on the distances between the holes
is one that would move the measured attenuation length in the same direction
at all measurement positions, so we do not include them in the fit.  
The uncertainties on the distances between the holes 
are small compared to the remaining uncertainties.
If we vary the distances by their uncertainties and refit,
the attenuation lengths change at most by $\pm2$\%.

\begin{figure}
    \epsfig{figure=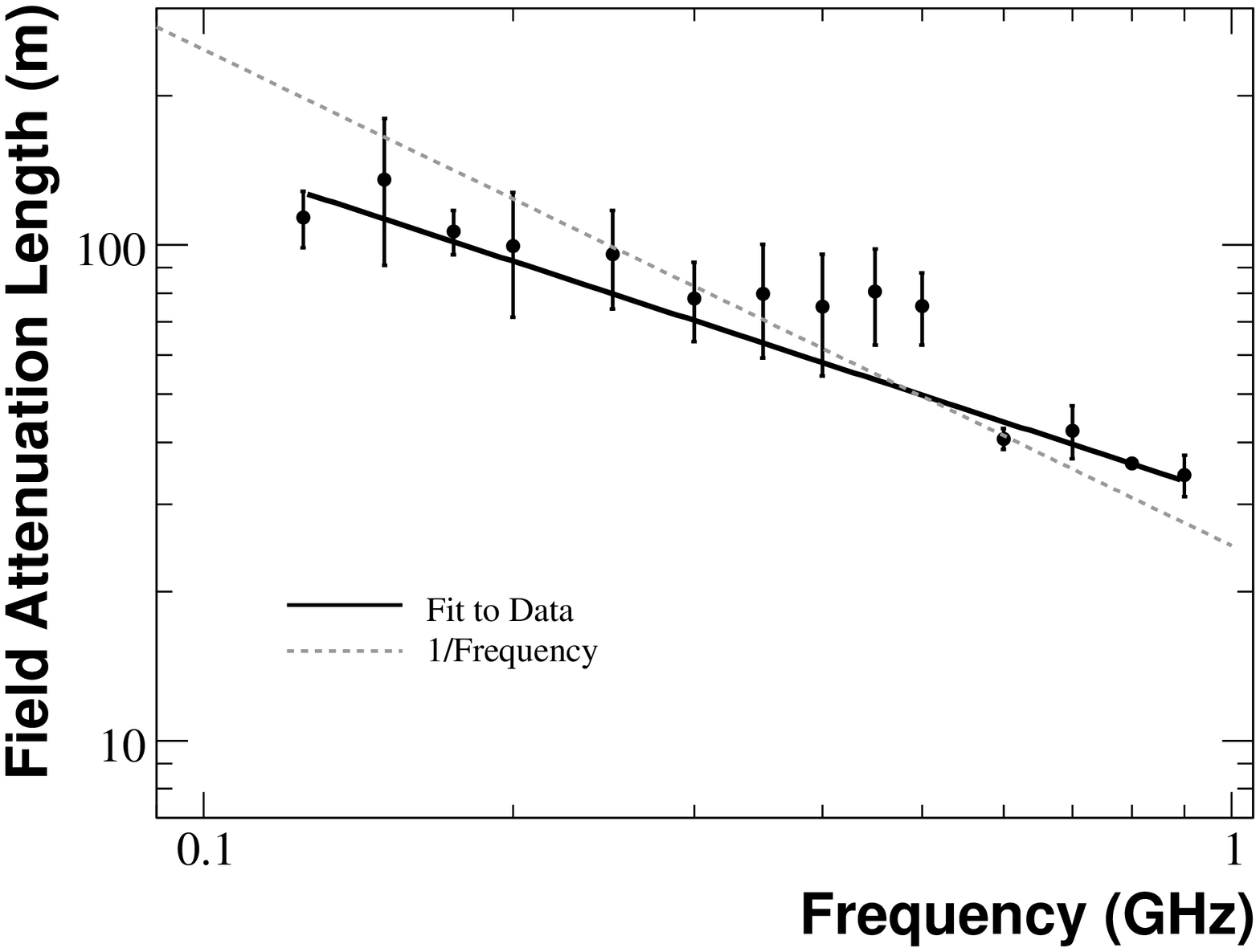, height=4.0in}
    \caption{ }
    \label{fig:avg}
  \end{figure}

We have also averaged the attenuation length at each frequency over all 
depths 
of 50~ft. and deeper and over all antenna types, 
and plotted the results in Figure~\ref{fig:avg}.  
This figure also shows the best fit power law function, with best fit values
$a=31.1\pm0.3$~m and $b=-0.68\pm0.04$ with a $\chi^2$/dof=13.3/12.  
For this plot the uncertainties only include the 
scatter between measured attenuation lengths using different antennas and at different depths in the same frequency bin.  
  
The attenuation lengths that we measured on this trip are 
significantly longer than those 
that we measured on our previous trip to the same mine.  
The measurements that we report here 
were made in boreholes that were drilled into the floor 
of the lowest level of the mine, whereas 
the previous measurements were made with holes drilled into the ceiling
and against the walls of
  the corridors.  Because the method  
  of mining consists of cutting under the salt and then blasting the wall above the cut, any 
  fractures that occur due to the process would tend to propagate upward.  
  This means that the salt 
  in the floor of the lowest level of the mine would tend to be less fractured and probably 
  exhibit less loss in the radio regime due to scattering.  

  \subsection{Index of Refraction}
  We also calculate the index of refraction of the salt using the direct transmission
  measurements that we made.  The index of refraction is defined as $n=\sqrt{\epsilon^\prime}$, 
  where $\epsilon^\prime$ is the real part of the dielectric permittivity.  We calculate 
  the speed of transmission through the medium using the distance between Boreholes~1~and~2 and 
  the time of travel of the signal pulse through the salt.  The time of travel was measured 
  by taking the difference between the time that the pulse was generated and the signal was 
  received and subtracting the known system delay.
  
\begin{figure}
    \epsfig{figure=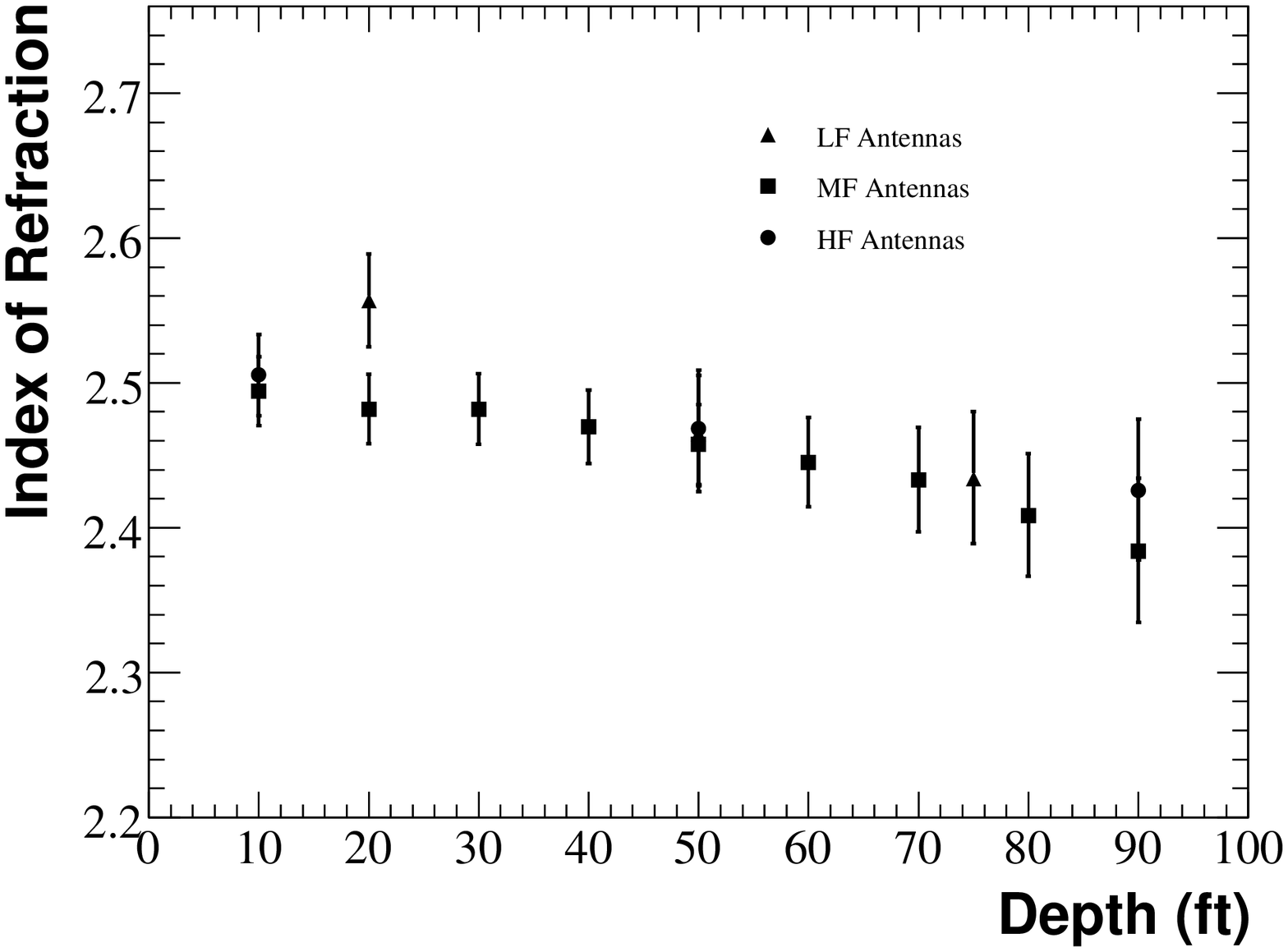, height=4.0in}
    \caption{ }
    \label{fig:n}
  \end{figure}

  Figure~\ref{fig:n} shows the measured index of refraction at each depth.  The index of refraction 
  that we measure is consistent with $n=2.45$, the index of refraction of rock salt.  We estimate 
  an uncertainty of $\pm3$~ns on the absolute system timing and $\pm1$~ft. on the distance between 
  the holes.  We also include the same uncertainty 
  due to potential deviation of the boreholes from vertical that was described in
  Section~\ref{sec:error}, which contributed $<\pm3$~ft. to the uncertainty. 
  
  \subsection{Reflections}
  
  At nearly every measurement position we observed
  at least one clear ``reflected'' signal in addition to the 
  ``direct'' signal that traveled along the straight line between 
  the transmitting and receiving antennas.
  Reflected signals can be easily distinguished from direct signals
  from their time of arrival at the receiver.  They allow
  us to probe distances greater than that between our drilled holes,
  and to probe different salt regions as well.  However,
  little is known about the loss in power incurred at the reflection, and
  reflected signals are often transmitted and received by the antennas at
  oblique angles, where the antenna response is less well understood.
  
  With our antennas in Boreholes~1~and~2, the entirety 
  of the shortest path between the
  receiver and transmitter was below the corridor
  where we were working, so we expected to
  see reflections from that corridor.  We did observe signals that
  were consistent with this interpretation.
  The measured time differences
  between the direct and reflected signals while we were transmitting
  between Boreholes~1~and~2
  were within approximately 10~ns of the expected time
  difference at all depths.  These reflected pulses traversed as
  much as 256~ft. (78~m) of salt, and a discrepancy of 10~ns corresponds to
  approximately 4~ft. (1.2~m) in salt.  
  With the antennas at the greatest depths, 
Most of the power in these reflected signal were out of the band of the antenna.
  We conclude that this power was reflected off of the input of the antenna and radiated 
  from the feed cable which was acting as a long-wire antenna.  In addition, the expected 
  beam pattern of a long-wire antenna is consistent with the angle of emission seen 
  with these reflections.
  
  With our antennas in Boreholes~1~and~3, the specular path of a 
  reflection from the
  corridor level is not incident on a salt-air boundary, but nevertheless
  at depths greater than 50~ft. (where the reflections did not interfere
  with the direct signal)
  we did observe signals whose timing were consistent
  with reflections from the corridor level within 10~ns.  In addition,
  with the antennas in Boreholes~1~and~3 we observed signals at all
  depths 30~ft. and deeper consistent
  with having reflected from 
  170~ft. above the corridor where we stood.  The 1300~ft. level of the
  mine is of course nominally 200~ft. above the floor of the corridor.
  Therefore, we observed
  signals having traveled through as much as 624~ft. of salt
  when the
  MF antennas were 90~ft. and 150~ft. deep.
  
  \subsection{Beam pattern in salt}
  
  The deep holes provided to us by the Cote Blanche mine 
  presented a unique opportunity
  to measure the beam pattern of our antennas while completely submerged in
  salt, although due to the distances involved, we did not probe a 
  broad range of angles.  
With one MF antenna at 90~ft. depth in Borehole~1, we took
data with the second MF antenna in Borehole~2 and at depths of
20~ft., 50~ft., 70~ft. and 90 ft., and in Borehole~3 at depths of
90~ft., 130~ft. and 150~ft.  The waveform at 20~ft. depth contained
interference from a reflected signal, so that data was not
included in this measurement.
  The remaining data probed the antenna beam pattern at
  angles of -6.2$^{\circ}$, -4.1$^{\circ}$, 0$^{\circ}$, 7.0$^{\circ}$ and 
  13.8$^{\circ}$ with respect to the horizontal, defined to be positive
  when the antenna in Borehole~1 was deeper than the antenna in the other
  borehole.  After correcting for $1/r^2$ power 
  loss as well as salt attenuation loss based on our measured values,
  we plot the measured power at each position relative to that measured
  at 0$^{\circ}$ in the same borehole.
  We use the same uncertainties as described in Section~\ref{sec:error}.
  The solid line in the figure is the expected beam pattern for a half-wave
  dipole in air:
  \begin{equation}
    \frac{dP_{\mathrm{Rel}}}{d\Omega} \propto \left[ \frac{ \cos{( \frac{\pi}{2}\cdot \sin{\theta})}}{\cos{\theta}} \right]^{2}.
  \end{equation}
  \begin{figure}
    \epsfig{figure=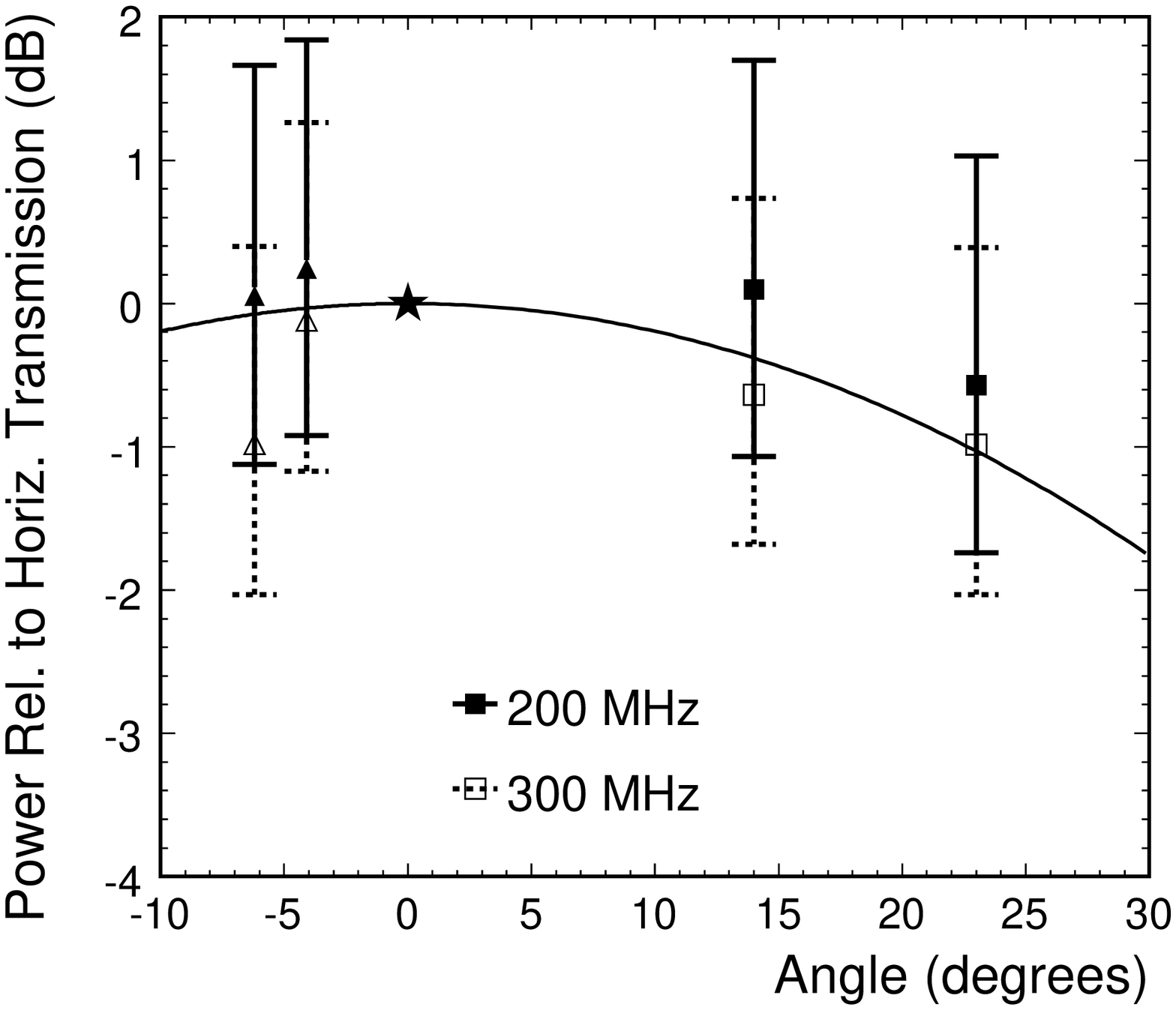, height=4.0in}
    \caption{ }
    \label{fig:beampattern}
  \end{figure}
  Although our uncertainties are large, we do not see a deviation from
  the half-wave dipole beam pattern.  The data is also not inconsistent with the peak of the
transmission being in the horizontal direction.
  
  \section{Review of Ground Penetrating Radar Data}
  \label{sec:gpr}
  Ground Penetrating Radar (GPR) measurements 
  were made in the Cote Blanche mine and other salt mines in the 
  1970's by Stewart and Unterberger ~\cite{s-unterberger,unterberger} to probe 
  discontinuities in the salt and the location of the
top of the dome.  They used a single frequency waveform generator at 440~MHz, a pair of 
  high-gain antennas that were pointed into the salt of interest, 
  and an oscilloscope to measure the time delay 
  between the transmitted signal and any received reflections off of discontinuities deep within salt.  
  To calculate the distance of the discontinuities, they used an index of refraction measured via 
  transmission across a pillar of salt (their result implies $n=2.62$).  
  
  They measured the location of the top of the Cote Blanche dome from within 
  the mine by transmitting the signal vertically through the ceiling.  Although the paper reports
that attenuation of 2-3~dB per 100~ft. is typical for radar in dry salt in general, 
which would imply field
attenuation lengths in the range 90-140~m,
at one measurement station they report a
  multiply-reflected signal that had traveled through a total 
path length of 4080~ft.~(1244~m) 
  and stated that it was the longest 
transmission distance observed in any mine.  
  We have compiled information in Table~\ref{tab:gprspecs} 
  from the discussion of the specifications of the system 
used to make the measurements in 
  References~\cite{s-unterberger}~and~\cite{unterberger}.
  
  The power received, $P_{Rx}$, is related to the power transmitted, $P_{Tx}$, by the Friis formula:
  \begin{equation}
    P_{Rx}=P_{Tx}\frac{G_{Tx} G_{Rx} \lambda^2}{(4\pi r)^2}
  \end{equation}
  where $G_{Tx}$ and $G_{Rx}$ are the gain of the transmitting and 
  receiving antennas, $\lambda$ is the wavelength of 
  the transmitted signal in salt, and $r$ is the distance between 
  the antennas.  Using the result of the Friis 
  formula together with the GPR system specifications, 
  we estimate that the minimum attenuation length allowed to detect the reflected signal 
  over the longest distance observed (see Table~\ref{tab:gprspecs}) is 
138~m, assuming 
  that they detected the signal just at the sensitivity threshold of 
their system.  
  
  This attenuation length is not inconsistent with 
some of our measurements
in the same frequency range, 
  although our results show generally higher losses.
Note that if there are variations in the clarity of the salt, 
the longest observed path length quoted in their paper would be more
likely to be from the clearest salt that they had sampled among all of their
measurements.
  We only sampled one section of salt 
with these measurements.

In order to replicate the low-loss results 
  of the GPR technique, any further measurements would have to be made with a portable 
  high-power system capable of making measurements at many locations.  

  We made a brief attempt at making measurements using the GPR technique during our visit to 
  the mine.  Using the same high-power pulser and oscilloscope, we used a pair of directional 
  antennas in an attempt to see reflections off of interfaces within the walls of the mine.  We did 
  see late reflected pulses, but we did not 
  understand that data well enough to conclude whether the reflections that we 
  saw came from within the salt, or were merely reflections down the corridor of the mine.

  \section{Simulation}

We have run a simulation to estimate the sensitivity of an array of antennas embedded in salt
  with the attenuation lengths that we have measured here.  
As much as possible, 
we use the same parameters as the simulated
  array described in the Hockley paper for direct comparison~\cite{hockley}.  
The simulated array consists
  of 10$\times$10$\times$10 dipole 
antennas with center frequency at 150~MHz and 50\% bandwidth
in a salt formation that is a cube 4~km on a side.   
  We require 4 antennas hit with a signal-to-noise ratio of 4$\sigma$.  
  The previous study in~\cite{hockley} considered a 300~m 
attenuation length at 300~MHz and 
a $1/\nu$ dependence, whereas
we have measured a $63\pm3$~m 
attenuation length at 300~MHz and the spectral index that we measure
is $-0.57\pm0.06$.
Reproducing the simulated Hockley array, we expect 
12 events per year compared to the order 10
events per year that they quote in their paper.
  Using the measured Cote Blanche attenuation lengths, we
 expect 1.4 events per year
  ultra-high energy neutrino flux~\cite{ess}, for a factor of 9 reduction in sensitivity
due to a lower measured attenuation length, but still a non-negligible rate over a one year
run time.

  \section{Conclusions}

We have measured field attenuation lengths in the frequency range 125-900~MHz using
broadband pulses transmitted and received by dipole antennas in
boreholes drilled 100~ft. into the floor of the 1500~ft. deep level of the the Cote Blanche 
salt mine.   The best fit power law that describes the
 measured frequency-dependent attenuation lengths is 
given by 
\begin{equation}
L_{\alpha}=\left(32\pm2\right)\cdot \left( \nu/1~\mathrm{GHz} \right)^{-0.57\pm0.06}
\end{equation}
with statistically dominated uncertainties.  This is the most precise measurement of
radio attenuation in a natural salt formation to date.

The main result of this paper is derived from data that we took on our third trip to the
Cote Blanche Salt Mine.  The data taken from the three trips indicates that the
room and pillar system of mining leads to regions of salt near the walls and ceilings 
of the corridors with short attenuation lengths ($\sim10$~m) which could be due to fracturing
brought about by the explosives used in that method.  We find that the 
salt that is just under the floor of a corridor does not suffer the same degradation,
which may be due to the cut along the floor that is made prior to the detonation of the
explosives to blast away the salt.

Although we expect there to be variation in salt properties between salt domes
as well as regions of salt within the same dome, we have compared our measurements
with previous measurements made at Cote Blanche and elsewhere.
Our result is not inconsistent with attenuation lengths measured at the Hockley Salt Mine.  We have attempted to compare
our measurements with the longest transmission distance reported by GPR
experts in the same mine.  
Our data generally showed greater attenuation than can be reconciled with that
result.  However, due to the variation observed in our data, 
some of our measurements do point to salt that could have been 
clear enough to be consistent.  That we strain to reconcile our numbers with the longest
transmission distance reported there may indicate that there are regions of the Cote Blanche mine
with salt that has less attenuation than in the regions where our measurements were made.

We have modeled a SalSA 
array of $10 \times 10 \times 10$ dipole antennas in boreholes that
reach 3~km depth to assess the sensitivity of a neutrino detector in salt with the attenuation
lengths that we report here to compare with the simulation study described in the Hockley attenuation
length paper.  
We find the lower best fit attenuation lengths measured at Cote Blanche reduce the sensitivity
of such a detector measured in terms
of event rate by a factor of 9, but still maintain an expectation of order one event per year
using a standard model for the expected ultra-high energy neutrino flux.  

In order for a next-generation neutrino
detector in salt that measures $\sim10$~GZK events/year to be feasible, attenuation
lengths longer than the $\sim100$~m lengths reported here 
will likely be necessary.  However, since this result is based on a limited
region of the dome, and past GPR results in the same mine 
may have pointed clearer salt, there are reasons to be
optimistic that there is clearer salt in other locations 
whose radio loss
 has not yet been precisely measured.  For this to be pursued further, we believe that
developing a portable radar system is a 
compelling next step so that the radio loss of the salt can
be sampled in
many locations on a single visit to a mine.

\section{Acknowledgements}
We are grateful to
 mine manager Gordon Bull at 
the North American Salt Company for supporting our project and 
his generosity in permitting us to
carry out this work in the Cote Blanche mine on three separate visits.  We
would also like to thank Scott Fountain, Robert Romero and the 
other miners at Cote Blanche that provided us with
 constant assistance on our visits to make these
measurements a success,
and for drilling custom boreholes  
for us which make these measurements unique in the world.
We would also like to thank the High Energy Physics Division of the 
U.S. Department of Energy,
the U.K. Particle Physics and Astronomy Research Council and
the Royal Society for funding this project.

  

\newpage
Figure~\ref{fig:saltmap}.  Map showing section of 1500 ft. level of the mine where we made 
our measurements at Boreholes~1,~2,~and~3. The letters and numbers follow the naming system 
for the corridors used by the mine.  The hatched regions are the salt surrounding the corridors.

Figure~\ref{fig:systemdiagram}.  System diagram.

Figure~\ref{fig:timedomain12_13}.  Pulses that propagated directly between the transmitter 
and receiver between Boreholes~1~and~2 between Boreholes~1~and~3 using the MF antennas at 
a depth of 90~ft.  We plot voltage multiplied by the distance between transmitter and 
receiver for each pulse, which should be constant in the absence of any attenuation.

Figure~\ref{fig:freqdomain12_13}.  Fourier transforms of pulses in Figure~\ref{fig:timedomain12_13}.

Figure~\ref{fig:s21}.  Measured transmission for the custom-made MF antennas.

Figure~\ref{fig:depth}.  Attenuation length at 250~MHz measured with the MF antennas.  
The uncertainties are smaller for shorter attenuation lengths since variations in voltage 
have less of an impact when greater loss is observed.

Figure~\ref{fig:freq}.  Attenuation length shown as a function of frequency for two depths 
and all three pairs of dipoles.

Figure~\ref{fig:avg}.  Average attenuation length, shown as a function of frequency.  
The uncertainties shown in this plot are the scatter of the measured attenuation lengths 
using different antennas and at different depths.

Figure~\ref{fig:n}.  Index of refraction measured at each depth.

Figure~\ref{fig:beampattern}.  Measurement of beam pattern while one MF antenna was at 90 ft. 
depth in Borehole~1 and the other was at various depths in Borehole~2~or~3.  The square markers 
are measurements taken with the second MF antenna in Borehole~2 and the triangular markers with 
the second MF antenna in Borehole~3.  The star denotes 0~degrees, where the attenuation is by 
definition 0~dB.

\newpage
\begin{table}
    \caption{Measured attenuation lengths at a few frequencies of 
interest based on fitting the data at 50~ft. and 90~ft. depths
 to the power law function as described in the text. }
   \begin{center}
      \begin{tabular}{cc} \hline

Frequency  & Measured  Attenuation Length \\ 
(MHz) & (meters) \\ \hline 
150 & 93 $\pm$ 7  \\
300 & 63 $\pm$ 3 \\
440 & 51 $\pm$ 2 \\
900 & 36 $\pm$ 2\\ \hline
      \end{tabular}
    \end{center}
    \label{tab:attnlengths}
\end{table}

\begin{table}
  \caption{Specifications of the GPR system 
    used in~\cite{s-unterberger}~and~\cite{unterberger}, 
    and an estimation of the minimum attenuation 
    length needed to see the longest path length 
    reflection observed.}
  \begin{center}
    \begin{tabular}{lcc}
      \multicolumn{3}{c}{}{} \\ \hline
      \multicolumn{3}{c}{System Specifications}\\\hline
	Antenna Gain & \multicolumn{2}{c}{17 dB}  \\
	Frequency & \multicolumn{2}{c}{440 MHz}  \\
	Maximum Transmission Distance & \multicolumn{2}{c}{1244 m} \\ \hline 
	\multicolumn{3}{c}{Attenuation Length Estimation}  \\\hline
	Peak Power Output & 40 dBm & 10 W\\
	Loss from transmission (from Friis formula) & 61 dB & $1.3 \times 10^{6}$ \\
	Power received (assuming no attenuation) & -21 dBm & $8.0\times 10^{-6}$ W \\
	System Sensitivity & -100 dBm & $10^{-13}$ W\\
	Maximum attenuation allowed & -79 dB & $1.3\times 10^{-8}$\\
	Attenuation length & \multicolumn{2}{c}{$>138$~m}\\ \hline 
      \end{tabular}
  \end{center}
  \label{tab:gprspecs}
\end{table}
\end{document}